\documentclass[10pt,twocolumn]{article}
\usepackage{lmodern}
\usepackage{algorithm}
\usepackage[noend]{algorithmic}
\usepackage{booktabs}
\usepackage[font=sf]{caption}
\usepackage[margin=1in]{geometry}
\usepackage{graphicx}
\usepackage{microtype}
\usepackage[nospace,superscript,compress]{cite}
\usepackage[parfill]{parskip}
\usepackage{setspace}

\title{\textsf{\huge Privacy Preserving Analytics on Distributed Medical Data}}

\makeatletter
\let\@fnsymbol\@arabic
\makeatother

\author{Marina Blanton\footnote{\sf Department of Computer Science and Engineering, University at Buffalo}, Ah Reum Kang\footnotemark[1], Subhadeep Karan\footnotemark[1], Jaroslaw Zola\footnotemark[1] $^{,}$\footnote{\sf Department of Biomedical Informatics, University at Buffalo}}

\date{}

\newcommand{\calX}{\mathcal{X}}

  {\centering
  \begin{minipage}{#1}
  \begin{algorithm}[H]}%
  {\end{algorithm}
  \end{minipage}\par
\vspace{\belowdisplayskip}}

\newcommand{\circleit}[1]{{\textcircled{\scriptsize \textsf{#1}}}}

\begin{document}
\maketitle

\begin{abstract}
\textbf{Objective:} To enable privacy-preserving learning of high quality generative and discriminative machine learning models from distributed electronic health~records.

\textbf{Methods and Results:} We describe general and scalable strategy to build machine learning models in a provably privacy-preserving way. Compared to the standard approaches using, e.g., differential privacy, our method does not require alteration of the input biomedical data, works with completely or partially distributed datasets, and is resilient as long as the majority of the sites participating in data processing are trusted to not collude. We show how the proposed strategy can be applied on distributed medical records to solve the variables assignment problem, the key task in exact feature selection and Bayesian networks learning. 

\textbf{Conclusions:} Our proposed architecture can be used by health care organizations, spanning providers, insurers, researchers and computational service providers, to build robust and high quality predictive models in cases where distributed data has to be combined without being disclosed, altered or otherwise compromised.
\end{abstract}


\section{Introduction}

Machine Learning (ML) is well accepted for studying and using biomedical data, with a promise of personalized, predictive and preventive medicine\cite{Kononenko2001,Krumholz2014}. ML methods are increasingly integrated into modern health informatics solutions, including clinical decision support systems\cite{Shin2006}, clinical trial design tools\cite{Lipkovich2017} and telemedicine platforms\cite{Turvey2017}. However, to be reliable and effective, these methods often require exact (i.e. globally optimal) algorithms, and significant amounts of input data to learn from. This is challenging because the volume and variety of the data that can be accumulated in health records of a single institution, for example a providers network, is intrinsically limited (consider for example rare diseases\cite{MacLeod2016}). At the same time, in both clinical and research setups, sharing of the biomedical data across institutional boundaries is heavily guarded by privacy considerations, with corresponding regulatory policies and guidelines, e.g., HIPAA in USA\cite{Baumer2000}. Moreover, often times organizations are simply unwilling to grant access to their data, due to concerns about competitive advantage or liability. Consequently, the available distributed data remains vastly underutilized\cite{Steinberg2009}, as the interested parties struggle to securely and effectively integrate~it.

The common approach in such situations is to either seek consent to release and share the data, anonymize the data\cite{Benitez2010}, or use some privacy-preserving techniques, e.g., differential privacy\cite{dwork2006}. However, seeking consent in many cases is impossible, especially for archival data, and anonymization typically requires retaining significant and valuable portions of the data. Consequently, these approaches are neither scalable nor sustainable. These issues are largely addressed by differential privacy methods. However, because differential privacy in its fabric involves data alteration, e.g., by adding noise, it may not be suitable for applications where high quality models are desired.

Here we propose an alternative architecture for privacy-preserving computations on distributed EHR data, which is free of the above limitations. Our architecture does not require alteration of the input data, is suitable for processing completely or partially distributed datasets, and is guaranteed to maintain data privacy as long as the majority of the sites processing the data are trusted to not collude. We demonstrate how, without ever disclosing distributed EHR data used in the underlying computations, this platform can be used to execute variables assignment problem, a key component to feature selection or to building exact ML models such as Bayesian networks.





\section{Proposed Architecture}

\begin{figure*}
\centering
\includegraphics[scale=0.25]{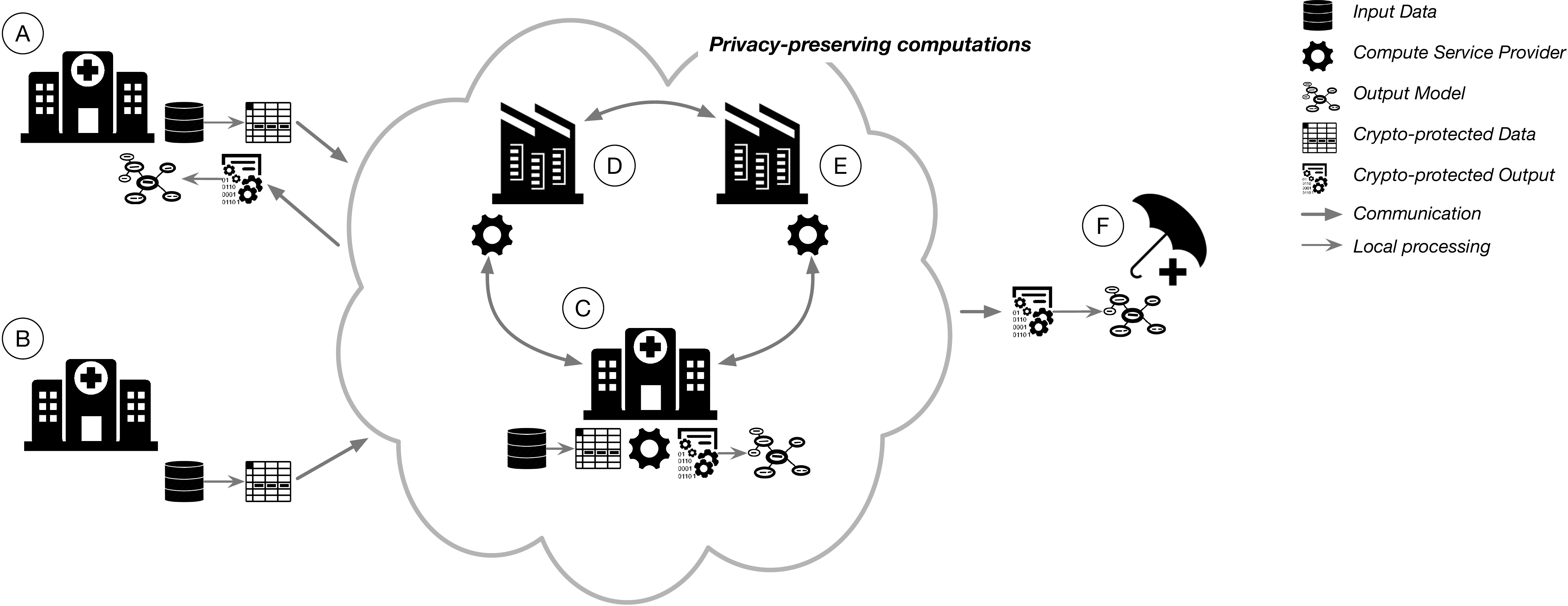}
\caption{Schematic representation of our proposed architecture for privacy-preserving computations.}\label{fig:overview}
 \end{figure*}

Our proposed architecture for privacy-preserving computations is outlined in Figure~\ref{fig:overview}. We divide all participants into three groups: the data owners (DOs), computational service providers (CSPs), and output recipients (ORs). We put no constraints on how these three groups are formed, and a single entity can be involved in a protocol by assuming on one or more of the above roles. The data owners (e.g., hospitals \circleit{A}, \circleit{B}, and \circleit{C}) locally pre-process their privately held data and securely enter them into the computation. The computational service providers (e.g., entities \circleit{C}, \circleit{D},~\circleit{E}) carry out the computation on cryptographically-protected data, on behalf of other participants. Upon completion of the computation, they communicate cryptographically-protected result of the computation to ORs  (e.g., hospital \circleit{C} and insurance company \circleit{F}), who locally reconstruct the output and learn the result.

It is important to understand what it means to ``securely enter data into the computation'' and to ``carry out the computation on cryptographically-protected data.'' Specifically, computing on cryptographically-protected data means that no CSP can understand or learn information about the data it receives and uses in the computation. This is because all computation is carried out using semantically secure encryption, which does not allow observers to deduce any patterns or other information about the encrypted data. Yet, the encryption has special properties that make it possible to compute directly on the protected data. Similarly, securely entering private data into the computation means that each DO first applies cryptographic protection to its data, and then communicates it to the appropriate CSPs, where each CSP cannot remove the protection after receiving~the~data.


The above setup is flexible enough to naturally fit several broad categories of collaborative and individual computing needs.  In particular, a number of parties with a private input each can engage in secure function evaluation themselves and learn the result (or their respective results). Alternatively, they can choose a subset of them, a number of outside parties, or a combination of the above to carry out the computation, while each data owner distributes their private data to the parties who carry out the computation. Another very important scenario involves a single entity outsourcing its computation to a number of computational service providers. In this case, that entity will be the only data owner and output recipient and all other parties learn nothing. 

As we mentioned earlier, our key requirement is that the CSPs learn nothing about the data they are processing (i.e., there is a mathematical proof that it is not possible). We also require that the ORs learn nothing about the original data other than the output of the agreed-upon computation that they receive. Lastly, any DO who is not an OR must also learn nothing. Note that it is possible to configure the computation in such a way that ORs learn different output from the computation, but it still must hold that no additional information is available to the ORs. These security notions can be rigorously specified using formal definitions from cryptography (see for example\cite{goldreich09}), however, providing such definitions and the corresponding security proofs is beyond the scope of this article.

Because proper protection of the private data throughout the computation is our primary goal, it is important to spell out the assumptions with respect to this setup to ensure that no information leakage is possible. In a setup with multiple CSPs, it is assumed that a fraction of them will be honest and will not conspire with other CSPs in an attempt to learn unauthorized information about the data. That is, some CSPs can be corrupt and colluding, but there should be a sufficient number of other CSPs who are not. If we denote the total number of CSPs by $\beta$, there is a threshold $t < \beta$ that indicates how many corrupt and colluding CSPs are tolerated without compromising security guarantees of the solution. For that reason, it is important to choose CSPs in a way that they are not expected to be simultaneously corrupt and colluding. For example, one CSP can be maintained by a professional association and others can come from competing health networks or cloud service providers. It is common that $t < \beta/2$.

While the general architecture described above, and demonstrated in Figure~\ref{fig:overview}, allows for fully secure solutions that provably protect all information about private data, in some cases it is beneficial to employ more specific configurations to reduce the overhead of privacy-preserving computation. For the problem we are addressing, we see two possible configurations with respect to how the problem is setup and what entities comprise the different categories of participants. In particular, in the first setting, a number of institutions, each with an insufficient amount of data to draw robust conclusions about the data set independently, combine their data sets in privacy-preserving computation and perform analysis of the joint data set. The result of the computation (such as the features of the combined data set) are available to all institutions who contributed their data sets. In the second case, one institution with insufficient amount of data seeks help of other organizations in analyzing medical data. These other organizations also become data owners and contribute their data to the computation, but the result of the computation is revealed only to the originating party. Because the first setup is of a great importance and allows for very significant computational savings compared to the general solutions, we are going to assume this problem formulation in the rest of this work. Furthermore, we make an additional assumption that the set of CSPs is also composed entirely of the participants who learn the result of the computation. In other words, in this setting the sets of DOs and ORs are the same and a subset of them are chosen as CSPs to run the computation on behalf of the entire group.

\section{Example Application}

To demonstrate our proposed setup in the actual application, we consider the variables assignment problem, typically referred to as parents assignment problem\cite{Koivisto2006}. This problem is a critical component to Bayesian networks learning, Markov blankets identification, and in general, feature selection\cite{Koivisto2006,Karan2017}. In short, given a set of variables and a target variable, our task is to select those variables that best explain the target variable, based on the input data with the variables' observations, and some scoring criterion. We can use the resulting assignment in multiple ways, for example, to build classifiers where target variable becomes label, and its parents are predictors, or more general machine learning models, like Bayesian networks. We note that the parents assignment problem is frequently entangled into biomedical applications, especially in clinical decision support systems, where classifiers and probabilistic graphical models are directly constructed from~EHR.

\subsection{Problem Statement}

Formally, we define the problem as follows. We are given a set of $n$ variables $\calX = \{X_1, X_2, \ldots, X_n\}$, where each $X_i$ represents one feature of interest. The features, e.g., age, gender, BMI, medical diagnosis codes, etc., are selected with respect to the question we wish to model. For the selected features, we use EHR to construct a table $\mathcal{D}$ of size ${n\times m}$ consisting of $m$ instances, where each instance represents one patient (see toy example in Table~\ref{tab:data}). This table becomes the input data to evaluate function $s(X_i, Pa(X_i))$, where $Pa(X_i)$ represents the currently considered parent set of $X_i$. This function quantifies how well variables in $Pa(X_i) \subseteq \calX - \{X_i\}$ explain a target feature $X_i$. While there are many ways in which function $s$ can be constructed\cite{DeCampos2001}, in this work we focus on the commonly used MDL score\cite{Schwarz1978}:
\[
  s(X_i,Pa(X_i)) = \sum_{j}^{q_{Pa(X_i)}} \sum_{k}^{r_i} \left(N_{ijk} \times \log{\frac{N_{ij}}{N_{ijk}}}\right) + nc,
\]
with $nc = 0.5 \times q_{Pa(X_i)} \times \log(m) \times (r_i - 1)$. Here, $r_i$ is the number of states (or arity) of $X_i$, $\displaystyle q_{Pa(X_i)} = \prod_{X_j\in Pa(X_i)} r_j$ is the combined number of states that variables in $Pa(X_i)$ can assume, and $N_{ij}$ and $N_{ijk}$ are respectively the counts of instances in $\mathcal{D}$ such that variables in $Pa(X_i)$ are in state $j$, and the counts of instances such that variables in $Pa(X_i)$ are in state $j$ and $X_i$ is in state~$k$. For example, consider Table~\ref{tab:data}, and configuration in which $X_{i}$ denotes \textsf{T2D}, and $Pa(X_i)$ consists of \textsf{Sex} and \textsf{Age}. Then we have that $q_{Pa(X_i)} = 2 \times 2$, $N_{ij} = 2$ if $j$ represents assignment $\textsf{Sex} = \textsf{F}$ and $\textsf{Age}=\textsf{[18-45)}$, and if $k$ corresponds to the state $\textsf{T2D}=\textsf{1}$ then $N_{ijk} = 1$. Note that $N_{ij}$ and $N_{ijk}$ are dependent on $Pa(X_i)$, but since $Pa(X_i)$ is alway clear from the context we omit it in our notation.

Intuitively, MDL score uses~$\mathcal{D}$ to estimate how much information, expressed by entropy, $Pa(X_i)$ provides about $X_i$. To guard against overfitting, it penalizes, via term $nc$, models with too many variables. Having input data $\mathcal{D}$, scoring criterion~$s$, and feature of interest $X_i$, our task is to find set $Pa(X_i)$ for which $s(X_i,Pa(X_i))$~is~minimized. The resulting set is the set of parents of $X_i$.

\begin{table}
\caption{Example patient data with four features.}\label{tab:data}
\centering
\textsf{\small
\begin{tabular}{ccllc} 
\toprule
$\ell$ & \textbf{Sex} & \textbf{Age} & \textbf{Race} & \textbf{T2D} \\
\midrule
1 & F & [18-45) & White & 1 \\
1 & M & [18-45) & Asian & 0 \\
1 & F & [45-65) & White & 0 \\
\midrule
2 & M & [45-65) & Black & 1 \\
2 & F & [45-65) & White & 1 \\
2 & F & [18-45) & Black & 0 \\
\bottomrule
\end{tabular}
}
\end{table}

In this work, we are interested in a variant of the problem where data $\mathcal{D}$ is distributed among several organizations (i.e. data owners) that jointly wish to solve the parent assignment. While the organizations do not want to reveal their data, they cooperate to agree on the data representation, and on how computations will be performed. Specifically, the cooperating organizations decide on how many and which features should be included in~$\calX$, and how each feature $X_i$ should be represented (e.g., how it is encoded, its arity $r_i$, etc.). For example, in Table~\ref{tab:data} age is discretized in one particular way that must be respected by all data owners. At the same time, information about how many instances (e.g., patient records extracted from EHR) an organization enters into the computation remains protected. If we denote the number of data owners by $\alpha$, then the total number of instances in $\mathcal{D}$ can be represented as $\displaystyle m = \sum_{\ell=1}^{\alpha} m^{(\ell)}$, where $m^{(\ell)}$ is the undisclosed number of instances contributed by data owner $\ell$. For instance, in our example in Table~\ref{tab:data} we distinguish $\alpha = 2$ data owners, each contributing $m^{(1)} = m^{(2)} = 3$ data instances. We will denote the data set maintained by data owner $\ell$ by $\mathcal{D}^{(\ell)}$.





\subsection{Conventional Algorithm}

Before discussing our privacy-preserving approach to parents assignment on distributed $\mathcal{D}$, we first explain the standard approach. The general idea is for a given $X_i$ to consider and evaluate $s$ for all possible candidate parent sets of growing size, starting from empty set. In practice, it is advantageous to consider slightly extended version of the problem\cite{Karan2017}. Suppose that instead of selecting parents of $X_i$ from $\calX - \{X_i\}$, we consider only some subset $U \subseteq \calX - \{X_i\}$ of variables. If the set $U$ has the property that its score $s(X_i,U)$ is lower than the score of any of its subsets, that is $\displaystyle \forall_{U'\subset U}~~s(X_i,U) < s(X_i,U')$, we will call it a maximal parent set. In other words, if $U$ is a maximal parent set, then all variables in $U$ are optimal parents of $X_i$. It turns out that by computing and storing all maximal parent sets of $X_i$ in $\calX - \{X_i\}$ we can efficiently select optimal parents of $X_i$ from any subset of $\calX - \{X_i\}$. This property is very practical and directly applicable in efficient learning of Bayesian networks. Hence, in Algorithms~\ref{alg:entropy}~and~\ref{alg:mps} we summarize the procedure to enumerate all maximal parent sets for $X_i$.

{\small
\begin{algorithm}[t]
  \caption{\textsc{Entropy}} \label{alg:entropy}
  \begin{algorithmic}[1]
    \setstretch{0.975}

    \REQUIRE Variable $X_i$, set $U$

    \ENSURE Entropy $H(X_i|U)$
    
    \STATE $h \leftarrow \displaystyle\sum_{j}^{q_{U}} \sum_{k}^{r_i} \left(N_{ijk} \times \log{\frac{N_{ij}}{N_{ijk}}}\right)$
    \RETURN {$h$}
  \end{algorithmic}
\end{algorithm}
}
{\small
\begin{algorithm}[t]
  \caption{\textsc{MaximalParentSets}}\label{alg:mps}
  \begin{algorithmic}[1]
    \setstretch{0.975}

    \REQUIRE Variable $X_i$, threshold $l_{max}$    
    \ENSURE Maximal parent set structure $PG_i$

        \STATE $H =\textsc{\textrm{Entropy}}(X_i, \calX - \{X_i\})$ 
        \STATE $Q \leftarrow \emptyset$ \label{line:empty_start}
        \STATE $s \leftarrow \textsc{Entropy}(X_i, \emptyset)$
    	\STATE $PG_i.insert((s, \emptyset))$ \label{line:empty_end}
        \STATE $Q' \leftarrow \{\{ X_1 \}, \ldots, \{X_{i-1}\}, \{X_{i+1}\}, \ldots, \{X_n\}\}$
		\STATE $l \leftarrow 1$
        \WHILE {$(Q' \neq \emptyset) \wedge (l \leq l_{max})$}
        	\STATE $Q \leftarrow Q'$
            \STATE $Q' \leftarrow \emptyset$
            \STATE $B \leftarrow \emptyset$
        	\FOR {$U \in Q$}
             	\STATE $N \leftarrow \{ {U \cup \{X_j\} | X_j \in \calX - \{ X_i \} - U\}}$ 
                \STATE $nc \leftarrow 0.5 \times q_{U} \times \log(m) \times (r_i - 1)$ \label{line:mdl_start}
            	\STATE $s \leftarrow nc + \textsc{Entropy}(X_i, U)$ \label{line:mdl_end}
                \STATE $s' \leftarrow \textsc{\textrm{BestSubset}}(PG_i, U)$
                \STATE $w \leftarrow nc + H$ \label{line:w}
               
                \IF {$s' \leq w$}
                    \IF {$s < s'$} \label{line:pgi_start}
                    	\STATE $PG_i.insert((s,U))$
                   	\ENDIF \label{line:pgi_end}
                	\STATE $Q' \leftarrow Q' \cup N$
              	\ELSE
                    \STATE {$B \leftarrow B \cup N$} \label{line:b_start}
                \ENDIF
            \ENDFOR
            \IF {$B \neq \emptyset$}
            	\STATE $Q' \leftarrow Q' - B$
           	\ENDIF \label{line:b_end}
        	\STATE $l \leftarrow l + 1$
        \ENDWHILE
        \RETURN $PG_i$
  \end{algorithmic}
\end{algorithm}
}

Algorithm~\ref{alg:entropy} outlines a helper procedure to estimate conditional entropy of $X_i$ given some set of variables~$U$. The algorithm depends on counts $N_{ij}$ and $N_{ijk}$, which in the conventional approach are extracted from $\mathcal{D}$ using some fast counting method, for example\cite{Karan2018}. To perform enumeration, in Algorithm~\ref{alg:mps}, we organize all possible subsets of $\calX - \{X_i\}$ into a subset lattice. We explore the lattice by performing breadth first search traversal, starting from empty set (lines~\ref{line:empty_start}--\ref{line:empty_end}) and then considering subsets of increasing size. Here we use $Q$ to denote the list of subsets processed at the current layer of the lattice, and $Q'$ to denote the list of subsets that should be processed at the next layer. For each considered subset $U$, we evaluate scoring function $s$, in this case MDL (lines~\ref{line:mdl_start}--\ref{line:mdl_end}), which we next use to decide whether $U$ is a maximal parent set. Specifically, if $s$ improves over the lowest score $s'$ among all strict subsets of $U$, then $U$ is a maximal parent set, and hence we should retain it in the output structure $PG_i$ together with the score $s$ (lines~\ref{line:pgi_start}--\ref{line:pgi_end}).

To eliminate from consideration subsets that cannot be maximal parent sets, and hence reduce computational complexity of the traversal, we exploit theoretical bounds on MDL score. Without going into details, which are available in\cite{Tian2000,Karan2017}, we set the bound $w$ based on the minimal possible conditional entropy of $X_i$, and penalty term $nc$ (line~\ref{line:w}). If the considered subset $U$ cannot improve over the bound, none of its supersets (represented by $N$) can improve, and hence they should be removed from further consideration. Moreover, for a set in $N$ to be considered in the next layer, all it subsets must satisfy the bound $w$. We enforce this by maintaining set $B$ of all subsets that should not be processed in subsequent steps~(lines~\ref{line:b_start}--\ref{line:b_end}). Finally, we note that instead of considering all possible subsets, the entire enumeration process can be limited only to subsets with cardinality smaller than some predefined threshold $l_{max}$. This threshold can be selected based on $\mathcal{D}$, such that the entire algorithm remains exact, or it can be configured based on some prior information.

\section{Securing Computations}



Privacy-preserving computation over distributed data is often called secure (multi-party) computation in the security and cryptography literature. For that reason, we might use the term \textit{secure computation} to mean privacy-preserving computation.

When the dataset is distributed across multiple sites, it needs to be combined prior to being used in the computation. Because privacy-preserving computation on protected data always incurs higher costs than an equivalent computation on locally available unprotected data, we want to minimize the portion of the computation that operates on protected data. To lower the overhead of our solution, we employ two crucial optimizations. First, each data owner locally pre-processes its data prior to inputting it into the computation. This allows us to eliminate most of the expensive joint privacy-preserving computation on the combined data set. Second, the information that the participants will learn as part of the output can be opened, i.e. represented without cryptographic protection, as soon as it becomes available at an intermediate step of the computation. As this data does need to be protected, it can be used directly further speeding up computation. This idea was also the basis of optimizations in \cite{kerschbaum2011automatically} and in many cases was shown to have significant impact on performance.
In this work, we realize the first idea by having each data owner to locally pre-compute all $N_{ij}$s and $N_{ijk}$s using their locally available data sets. That is, data owner $\ell$ pre-computes $N_{ij}^{(\ell)}$s and $N_{ijk}^{(\ell)}$s using its local $\mathcal{D}^{(\ell)}$. The computation for determining the values of $N_{ij}$s and $N_{ijk}$s is expensive because it requires repeated access to the entire table $\mathcal{D}$. This means that if the parties instead enter their $\mathcal{D}^{(\ell)}$'s into the joint computation, compute combined $\mathcal{D}$ and consequently $N_{ij}$s and $N_{ijk}$s, the secure computation will incur a large runtime. However, if each DO locally computes $N_{ij}^{(\ell)}$s and $N_{ijk}^{(\ell)}$s and enters them into the joint computation, the combined values can be reconstructed inside secure computation very efficiently as $\displaystyle N_{ij} = \sum_{\ell=1}^\alpha N_{ij}^{(\ell)}$ and $\displaystyle N_{ijk} = \sum_{\ell=1}^\alpha N_{ijk}^{(\ell)}$. 

Before we proceed any further, let us identify portions of the algorithms that handle private data. Recall that the number of variables $n$ as well as the number of values that each variable $X_i$ can take, $r_i$, need to be agreed upon ahead of time and are public. The data itself, including the number of observations in each dataset, is private. This means that the number of observations in the combined data set, $m$, must be treated as private as well. To ensure the strongest possible data protection, any value that depends on a private data item needs to be considered private throughout the computation.

Consider Algorithm~\ref{alg:mps}. The input arguments, i.e. variable $X_i$ and threshold $l_{max}$, are open, while the content of table $\mathcal{D}$ must remain protected. The algorithm considers different possibilities for parent sets, and in the beginning the content of sets $Q$ and $Q'$ is known. 
The set $Q$ is subsequently updated based on the content of sets $Q'$ and $B$, the computation of which depends on private data. This is because the contents of $Q'$ and $B$ depend on the condition on line 17 that involves entropy $H$, which is private (i.e., computed using the data from $\mathcal{D}$). This means that the set of $U$ possibilities after the first iteration of the \textsf{while} loop needs to be treated as private and therefore all data (except $l$) in the remaining computation needs to be treated as private as well.

The above has significant implications on performance. For example, we need to execute the maximum number of \textsf{for} loop iterations to protect the size of $Q$ and hide the fact whether insertion takes place on line 19 by always inserting a record into $PG_i$ (real or fake). Furthermore, because the candidate parent sets $U$ tested by Algorithm~\ref{alg:mps} are now private, performance of Algorithm~\ref{alg:entropy} is also affected. In particular, $q_U$ and $r_i$ become protected, which requires the sums to be executed over the largest possible number of terms. In addition, the algorithm needs to privately retrieve the values of $N_{ij}$ and $N_{ijk}$ without revealing what values have been accessed (i.e., by touching all possible values in $\mathcal{D}$, or by employing more complex randomized techniques). This means that executing the algorithm on private data results in significant performance degradation. 


Now consider the setting where the computational parties are also output recipients, and thus they are entitled to observing the content of $PG_i$. Note that once it is determined that a parent set needs to be added to $PG_i$, it will remain in that data structure. Hence, once a pair $(s,U)$ is added to $PG_i$, it can be opened (because it will be part of the output), but the corresponding $s'$ should remain private.

If we would like to maintain the structure and efficiency of the algorithm, we need to know what sets $U$ proceed to the next algorithm iteration (stored in $Q'$).
If we reveal this information, this provides information about the outcomes of condition $s' \le w$. Then because the values of $s'$ and $nc$ might be guessable, the outcome of comparison $s' \le w$ reveals the lower bound of entropy $H$, which can be consequently used to narrow down the value of $H$, defined as the amount of uncertainly that variable $X_i$ contributes to the set.
Note, however, that this is a very limited amount of information that one can learn about the entire data set (and not about individual records that compose the dataset because this is a one-way irreversible computation). Furthermore, entropy computed on individual data sets $\mathcal{D}^{(\ell)}$ that the data owners contribute to the computation is expected to be similar to that of the combined set $\mathcal{D}$. Lastly, as the goal of the overall joint computation is to learn conditional dependencies between different variables, similar information is already a part of the output. Thus, this is an insignificant amount of leakage that does not reveal information about sensitive data and we allow it in order to maintain the structure of the computation and the algorithm's efficiency.

To improve performance of privacy-preserving computation, we use non-traditional implementation of certain functions. For example, the logarithm function is expensive to compute on private data and it is executed multiple times in each invocation of Algorithm~\ref{alg:entropy}. The division operation, and in general operators involving non-integer values, is non-trivial as well. Thus we replace the computation $\log(N_{ij}/N_{ijk})$ in Algorithm~\ref{alg:entropy} with $\log(N_{ij}) - \log(N_{ijk})$. Now note that each input in the logarithm function is a small integer between 0 and $m$. This means that we can pre-compute the logarithm function for all integers in the range and store them in an array. Then evaluating the logarithm function on private $N_{ij}$ or $N_{ijk}$ will amount to retrieving one value of the array at a private location. We devise an optimized solution for reading an element of an array at a private location for the purposes of this work, because of the frequency with which this function will be called in our algorithm.

\section{Discussion}

The ability to compute on distributed medical data is critical to advancing the use of ML techniques in health informatics. It is of special importance in cases like rare diseases, where the existing data is already very sparse, or when available data is imbalanced. With our platform, we make it possible to compute directly on distributed data, without ever exposing or modifying it. One way to think about the platform is as computing on encrypted data, where the encryption key is distributed (i.e. partitioned) between participants. This has the effect of completely preserving privacy of the patients described by the data, while maintaining the original utility of the data (as long as the participants do not collude to reconstruct the complete encryption key). Consequently, the platform can be directly used to build or train ML models that otherwise would be impossible to achieve. As the access to the data remains one of the major impediments in medical applications of ML\cite{zhang13}, the platform could of direct use to researchers and practitioners alike. 

Although we presented the platform in the context of one specific ML problem applied to categorized medical data, we note that the approach is generalizable. In fact, privacy-preserving computation is possible via generic techniques for any desired functionality, and specific constructs have been provided in the context of GWAS studies\cite{shahbazi16}, DNA sequences comparisons\cite{ayday14,zhang15}, and disease risk computation\cite{ayday13}, among others. However, we note that the transformation of standard algorithms into their privacy-preserving equivalents may be non-trivial, even if assisted by a dedicated compiler\cite{zhang13}. Moreover, the resulting privacy-preserving realization will be usually significantly slower than its standard counterpart. However, the platform is scalable in the number of participating sites, including both data owners and computational service providers. Specifically, the efficiency of the platform is typically not affected by the number of data owners, and thus can be easily expanded with potential new data sources. At the same time, the number of computational service providers may stay fixed, as long as it satisfies the threshold for the number of non-colluding providers. This has a practical implication for establishing a platform, since it is sufficient to include only a few computational service providers.

\bibliographystyle{unsrt}
\bibliography{jamia}
\end{document}